\documentclass{pasa}%

\title[DESAlert]{DESAlert: Enabling Real-Time Transient Follow-Up with Dark Energy Survey Data}
\author[A. Poci, K. Kuehn, et al.]{A.~Poci$^{1}$, K.~Kuehn$^{2}$\thanks{Corresponding Author, Email:kkuehn@aao.gov.au}, \and T.~Abbott$^{3}$, F.~B.~Abdalla$^{4}$, S.~Allam$^{5}$, A.H.~Bauer$^{6}$, A.~Benoit-L{\'e}vy$^{4}$, E.~Bertin$^{7}$, D.~Brooks$^{4}$, P.~J.~Brown$^{8}$, E.~Buckley-Geer$^{5}$, D.~L.~Burke$^{9,10}$, A.~Carnero Rosell$^{11,12}$, M.~Carrasco~Kind$^{13,14}$, R.~Covarrubias$^{14}$, L.~N.~da Costa$^{11,12}$, C.~B.~D'Andrea$^{15}$, D.~L.~DePoy$^{8}$, S.~Desai$^{16}$, J.~P.~Dietrich$^{16,17}$, C.~E Cunha$^{9}$, T.~F.~Eifler$^{18,19}$, J.~Estrada$^{5}$, A.~E.~Evrard$^{20}$, A.~Fausti Neto$^{11}$, D.~A.~Finley$^{5}$, B.~Flaugher$^{5}$, P.~Fosalba$^{6}$, J.~Frieman$^{5,21}$, D.~Gerdes$^{20}$, D.~Gruen$^{22,23}$, R.~A.~Gruendl$^{13,14}$, K.~Honscheid$^{24,25}$, D.~James$^{3}$, N.~Kuropatkin$^{5}$, O.~Lahav$^{4}$, T.~S.~Li$^{8}$, M.~March$^{18}$, J.~Marshall$^{8}$, K.~W.~Merritt$^{5}$, C.J.~Miller$^{26,20}$, R.~C.~Nichol$^{15}$, B.~Nord$^{5}$, R.~Ogando$^{11,12}$, A.~A.~Plazas$^{27,19}$, A.~K.~Romer$^{28}$, A.~Roodman$^{9,10}$, E.~S.~Rykoff$^{9,10}$, M.~Sako$^{18}$, E.~Sanchez$^{29}$, V.~Scarpine$^{5}$, M.~Schubnell$^{20}$, I.~Sevilla$^{29,13}$, C.~Smith$^{3}$, M.~Soares-Santos$^{5}$, F.~Sobreira$^{5,11}$, E.~Suchyta$^{24,25}$, M.~E.~C.~Swanson$^{14}$, G.~Tarle$^{20}$, J.~Thaler$^{30}$, R.~C.~Thomas$^{31}$, D.~Tucker$^{5}$, A.~R.~Walker$^{3}$, W.~Wester$^{5}$ (The DES Collaboration)\\
\affil{$^{1}$ Department of Physics and Astronomy, Macquarie University, NSW 2109, Australia}%
\affil{$^{2}$ Australian Astronomical Observatory, North Ryde, NSW 2113, Australia}%
\affil{$^{3}$ Cerro Tololo Inter-American Observatory, National Optical Astronomy Observatory, Casilla 603, La Serena, Chile}
\affil{$^{4}$ Department of Physics \& Astronomy, University College London, Gower Street, London, WC1E 6BT, UK}
\affil{$^{5}$ Fermi National Accelerator Laboratory, P. O. Box 500, Batavia, IL 60510, USA}
\affil{$^{6}$ Institut de Ci\`encies de l'Espai, IEEC-CSIC, Campus UAB, Facultat de Ci\`encies, Torre C5 par-2, 08193 Bellaterra, Barcelona, Spain}
\affil{$^{7}$ Institut d'Astrophysique de Paris, Univ. Pierre et Marie Curie \& CNRS UMR7095, F-75014 Paris, France}
\affil{$^{8}$ George P. and Cynthia Woods Mitchell Institute for Fundamental Physics and Astronomy, and Department of Physics and Astronomy, Texas A\&M University, College Station, TX 77843,  USA}
\affil{$^{9}$ Kavli Institute for Particle Astrophysics \& Cosmology, P. O. Box 2450, Stanford University, Stanford, CA 94305, USA}
\affil{$^{10}$ SLAC National Accelerator Laboratory, Menlo Park, CA 94025, USA}
\affil{$^{11}$ Laborat\'orio Interinstitucional de e-Astronomia - LIneA, Rua Gal. Jos\'e Cristino 77, Rio de Janeiro, RJ - 20921-400, Brazil}
\affil{$^{12}$ Observat\'orio Nacional, Rua Gal. Jos\'e Cristino 77, Rio de Janeiro, RJ - 20921-400, Brazil}
\affil{$^{13}$ Department of Astronomy, University of Illinois,1002 W. Green Street, Urbana, IL 61801, USA}
\affil{$^{14}$ National Center for Supercomputing Applications, 1205 West Clark St., Urbana, IL 61801, USA}
\affil{$^{15}$ Institute of Cosmology \& Gravitation, University of Portsmouth, Portsmouth, PO1 3FX, UK}
\affil{$^{16}$ Department of Physics, Ludwig-Maximilians-Universit\"at, Scheinerstr.\ 1, 81679 M\"unchen, Germany}
\affil{$^{17}$ Excellence Cluster Universe, Boltzmannstr.\ 2, 85748 Garching, Germany}
\affil{$^{18}$ Department of Physics and Astronomy, University of Pennsylvania, Philadelphia, PA 19104, USA}
\affil{$^{19}$ Jet Propulsion Laboratory, California Institute of Technology, 4800 Oak Grove Dr., Pasadena, CA 91109, USA}
\affil{$^{20}$ Department of Physics, University of Michigan, Ann Arbor, MI 48109, USA}
\affil{$^{21}$ Kavli Institute for Cosmological Physics, University of Chicago, Chicago, IL 60637, USA}
\affil{$^{22}$ Max Planck Institute for Extraterrestrial Physics, Giessenbachstrasse, 85748 Garching, Germany}
\affil{$^{23}$ University Observatory Munich, Scheinerstrasse 1, 81679 Munich, Germany}
\affil{$^{24}$ Center for Cosmology and Astro-Particle Physics, The Ohio State University, Columbus, OH 43210, USA}
\affil{$^{25}$ Department of Physics, The Ohio State University, Columbus, OH 43210, USA}
\affil{$^{26}$ Department of Astronomy, University of Michigan, Ann Arbor, MI 48109, USA}
\affil{$^{27}$ Brookhaven National Laboratory, Bldg 510, Upton, NY 11973, USA}
\affil{$^{28}$ Department of Physics and Astronomy, Pevensey Building, University of Sussex, Brighton, BN1 QH, UK}
\affil{$^{29}$ Centro de Investigaciones Energ\'eticas, Medioambientales y Tecnol\'ogicas (CIEMAT), Madrid, Spain}
\affil{$^{30}$ Department of Physics, University of Illinois, 1110 W. Green St., Urbana, IL 61801, USA}
\affil{$^{31}$ Lawrence Berkeley National Laboratory, 1 Cyclotron Road, Berkeley, CA 94720, USA}
}

\jid{PASA}
\doi{10.1017/pas.\the\year.xxx}
\jyear{\the\year}

\usepackage{rotating}
\usepackage{expdlist}
\usepackage{pdflscape}
\usepackage{longtable}

\begin{document}
\begin{abstract}
The Dark Energy Survey (DES) is currently undertaking an observational program imaging $1/4$ of the southern hemisphere sky with unprecedented photometric accuracy.  In the process of observing millions of faint stars and galaxies to constrain the parameters of the dark energy equation of state, the DES will obtain pre-discovery images of the regions surrounding an estimated 100 gamma-ray bursts (GRBs) over five years.  Once GRBs are detected by, e.g., the Swift satellite, the DES data will be extremely useful for follow-up observations by the transient astronomy community.  We describe a recently-commissioned suite of software that listens continuously for automated notices of GRB activity, collates useful information from archival DES data, and disseminates relevant data products back to the community in near-real-time.  Of particular importance are the opportunities that non-public DES data provide for relative photometry of the optical counterparts of GRBs, as well as for identifying key characteristics (e.g., photometric redshifts) of potential GRB host galaxies.  We provide the functional details of the DESAlert software, as well as the data products that it produces, and we show sample results from the application of DESAlert to numerous previously-detected GRBs, including the possible identification of several heretofore unknown GRB hosts.
\end{abstract}
\begin{keywords}
catalogs -- gamma ray burst: general -- methods: observational -- surveys -- virtual observatory tools
\end{keywords}
\maketitle%
\section{INTRODUCTION}
\label{sec:intro}

\begin{figure}
\begin{center}
\includegraphics[width=\columnwidth]{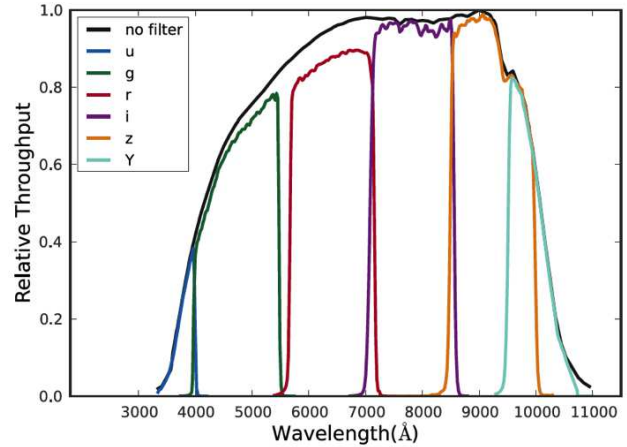}
\caption{Throughput as a function of wavelength of the DECam optical train, including the various filters the DES uses.  These throughputs are calculated relative to the use of no filter at 9000 ${\AA}$.}\label{Fig1}
\end{center}
\end{figure}

The Dark Energy Survey (DES) is an observational program covering 5000 square degrees of the southern sky, utilising the DECam instrument \cite{DECam} on the Blanco 4m telescope at Cerro Tololo Interamerican Observatory, from August 2013 to February 2018 \cite{DES}. Over the 525 nights of the survey, the DES will observe in five filters broadly similar to the SDSS griz filter set \cite{Gunn}, but with some important differences -- particularly, higher quantum efficiency at near-infrared wavelengths and the additional Y filter (see Figure \ref{Fig1}).  DES will reach a photometric accuracy of 1-2$\%$ for its two interleaved surveys -- the wide-field survey covering the full footprint and the supernova survey covering smaller regions with increased cadence \cite{Year1} -- and it will have significant overlap with other wide-area surveys, such as the Sloan Digital Sky Survey's Stripe 82 \cite{Sloan}, the Vista Hemisphere Survey \cite{VHS}, and the South Pole Telescope Survey \cite{SPT}.  The four Key Science programs of the DES comprise observations of SNe Ia, large-scale galaxy clustering, galaxy clusters, and weak gravitational lensing; together these four probes will be used to measure the Dark Energy Equation of State with unprecedented precision.  The DES data have many uses beyond these primary science goals, however.  In this work, we describe a service to provide data products to the transient observational community, particularly related to gamma-ray bursts located within the DES footprint.

Once a region of the sky has been observed by the DES, those observations will be useful whenever a transient (e.g., a gamma-ray burst) is detected in the same region. The DESAlert system is modeled upon the SkyAlert system \cite{Skyalert}, and it bears similarities to the SDSS transient notification system \cite{Cool}, though it focuses exclusively on data produced by the DES, and (at least initially) it relies on a single source for triggers.  When a VOEvent notice \cite{IVOA} is disseminated based on data from the Burst Alert Telescope (BAT) or X-Ray Telescope (XRT) aboard the Swift satellite \cite{Swift}, the DESAlert system parses the Notice for temporal and positional information, and then searches the DES data archives to find all observations of that region. \mbox{DESAlert} then provides finder images of the region derived from DES data, as well as a subset of data derived from DES observations of all nearby stars and galaxies.  The finder images show other objects near the GRB, while the catalog of stars provides nearby standards for the purpose of immediate relative photometry. Meanwhile, the galaxy catalog provides critical information on potential host galaxies -- especially magnitude and photometric redshift -- for the given GRB. All of these data products will be of use to those who need to make decisions regarding the allocation of scarce resources for follow-up observations of these GRBs.  For example, if a potential host has an extremely low (or extremely high) redshift, observers may decide that the GRB warrants further study; thus they would begin follow-up observations as soon as possible.  Alternately, if a host galaxy has unusual colors (as described by the multi-band DES observations), that may be indicative of unusual metallicity or dust content, again prompting observers to allocate observational resources to follow up the detection of these transients.

\begin{figure}
\begin{center}
\includegraphics[width=\columnwidth]{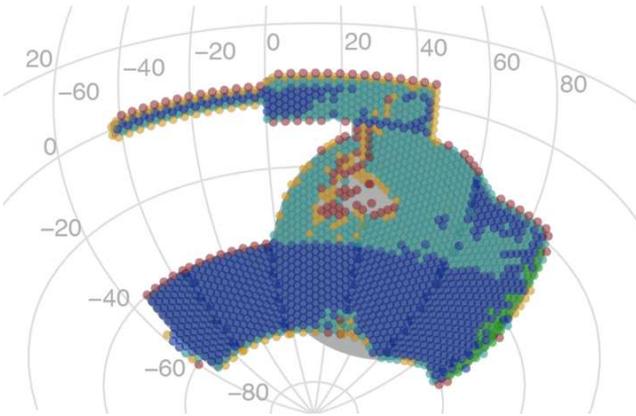}
\caption{The DES footprint, with current coverage in the i band shown.  Grey is the full survey area, while areas covered by 1, 2, 3, 4, or 5+ observations are colored red, orange, light blue, dark blue, and green, respectively.}\label{Fig2}
\end{center}
\end{figure}

By the conclusion of Year 2 in February 2015, DES had observed nearly all of its survey footprint, with multi-epoch imagery covering the vast majority of that area -- see Figure \ref{Fig2}.  Year 3 saw the completion of coverage of the entire footprint, with subsequent data-taking increasing the number of observed epochs (and thus the effective co-add survey depth) at each position within the footprint.  Given the area covered, and assuming randomly-distributed GRBs detected by Swift at a rate of ${\sim}$100 per year, we expect 10-20 GRBs annually to have DES pre-discovery images that are amenable to analysis and publication via the DESAlert algorithm.  We encourage all interested observers to take advantage of these data products provided to the astronomical community.

\section{The DESAlert Algorithm}
\label{sec:algorithm}

\begin{figure}
\begin{center}
\includegraphics[width=\columnwidth]{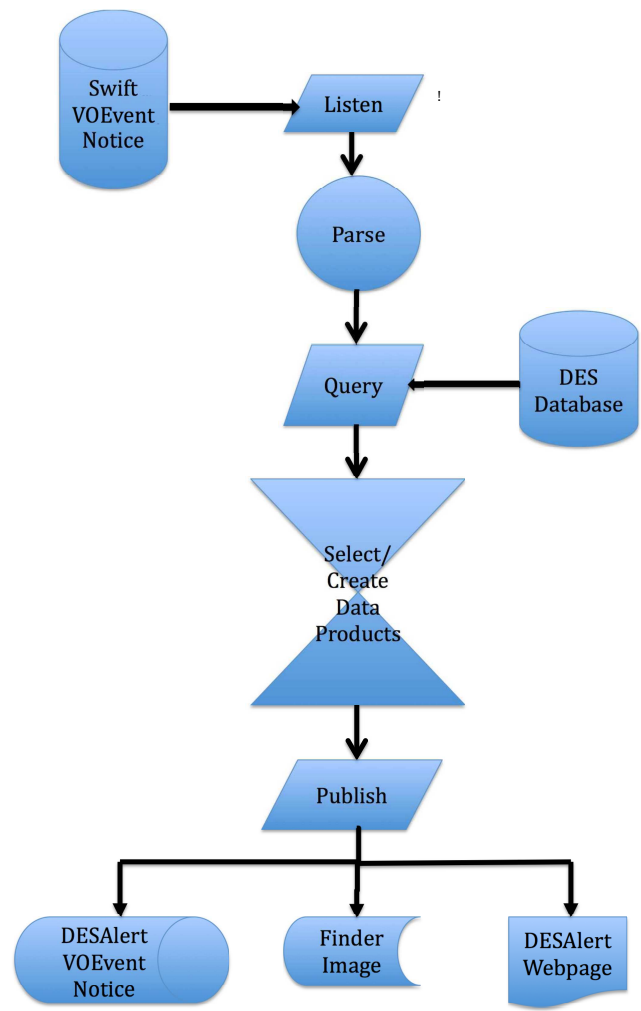}
\caption{The DESAlert algorithm flowchart}\label{Fig5}
\end{center}
\end{figure}

The DESAlert algorithm is shown schematically in Figure \ref{Fig5}.  Though it is similar in effect to other VOEvent-focused software such as Dakota or Comet \cite{Comet}, in order to maintain minimal dependency on external code, it does not rely on these software packages -- it does, however, makes use of the \mbox{voeventlib} libraries \cite{VOEventLib}.  The functional code for DESAlert is written in Python and SQL, and will be made available online via the Astrophysics Source Code Library \cite{ASCL}.  It encapsulates all functions necessary to complete the following steps:
\begin{itemize}
\item Listen for VOEvent Notices
\item Select events confirmed to be GRBs by Swift
\item Parse temporal and position data from Notices
\item Query the DES database for objects and images in a pre-determined region surrounding the GRB
\item Extract relevant archival DES data for stars and galaxies
\item Derive data products from extracted data 
\item Publish DESAlert data products as VOEvent Notices, as well as to the DESAlert webpage
\end{itemize}

First, the software opens a socket connection and receives (XML-formatted) messages from the VOEvent server as a series of packets, built up byte-by-byte until the full message is received.  Error checking ensures that malformed (or improperly received) messages are deleted before the algorithm returns to listen mode.  The XML is then parsed to determine the type of message (e.g., imalive, test, or observation).  The first of these is necessary for the socket connection to be maintained at the client end, and it results in an identical reply from the client (``imalive'') so that the socket connection is likewise maintained at the server end.  All other types except observation are discarded.  Observations, however, have relevant data for each GRB extracted, including GRB position (RA, Dec), burst time, and detecting instrument.  In order to ensure near-real-time response from DESAlert to initial GRB detections (almost always by the BAT instrument), we process the initial Notice for each GRB.  Typically, subsequent VOEvent notices from Swift's XRT and/or UVOT instruments refine the position (or position uncertainty) of the GRB; in that case we process the first such subsequent Notice in the same fashion as the initial notice, with new and improved information superceding the old.  In general, the XRT or UVOT position is sufficiently accurate (with uncertainties of only a few arcseconds) that processing of further Notices is not required, so they are discarded.

Based on the extracted GRB data, the python code then calls a custom-made jython-based command-line database interface to query the DES archival catalog of sources in order to find any stars or galaxies within a box ${\pm}$60 arcseconds in RA and Dec of the BAT-determined GRB position, or within a ${\pm}$30 arcsecond box for XRT-determined position.  These objects have previously been extracted from the processed (and coadded) DES Multi-Extension FITS images, and have had a great many parameters determined by the Data Management pipeline \cite{DESDM} running Source Extractor \cite{SEx}, including magnitudes (and uncertainties) for every observed filter, as well as object classification (either star or galaxy). 
%Once the algorithm ingests all objects in the large search area, it selects a smaller region centered on the GRB that is 60 arcseconds (plus 90$\%$ uncertainties in the GRB position) on a side, and finds all objects that have any portion within this region.  This is especially important for extended objects like galaxies, since GRBs are properly associated with galaxy hosts even when their positions correspond to the outer fringes of the object.
Once the appropriate data are acquired from the DES database they are then stored, via SQL Insert commands, in a local database connected to the machine running the DESAlert python code, and are made available to the public in several ways, as detailed in Section \ref{sec:products}.  A second query of the DES database finds all $\it{g}$, $\it{r}$, and $\it{i}$ images containing the GRB position, from which the nearby objects have been extracted.  

While access to this type of data via publicly-available online catalogs may be routine for GRB follow-up efforts, DESAlert has the advantage of accessing recent (i.e. proprietary) DES data, which includes objects with better than 2\% photometric accuracy down to magnitudes fainter than $\it{g}$=25.  This is significantly fainter than the limits of other catalogs, e.g., the Digitized Sky Survey or 2MASS \cite{2MASS}, often used for this purpose.  The DESAlert magnitude limits also compare favorably to those of GROND \cite{GROND}, while the expected photometric accuracy of DES is better, and DES also makes use of Y-band imagery.

\section{DESAlert Data Products}
\label{sec:products}

Once the software implementing the DESAlert algorithm has selected the relevant image segments and composed its catalog of nearby objects, it disseminates the information of potential relevance to those seeking to follow up the GRB detection with further observations.  The primary data products produced by DESAlert are based on pre-discovery images of the regions around Swift-detected GRBs.  The initial products are (XML-formatted) VOEvent Notices, listing:
\begin{itemize}
\item Basic GRB parameters, including position, time, and discovering instrument
\item Closest stars to the GRB, including positions and magnitudes
\item Closest galaxies to the GRB, including positions, magnitudes, photometric redshifts, and chance alignment probabilities
\item Links to the finder images (in .fz and .jpg formats)
\item Links to the DESAlert webpage containing a database with additional information for all objects near the GRB
\end{itemize}

Finder images provided by DESAlert are processed and coadded DES images, each of which covers an area on the sky of approximately 0.75 degrees on a side -- although depending on the position of the GRB and its associated uncertainty, several images may be provided to completely cover the nominal search area.  These images are provided in $\it{g}$, $\it{r}$, and $\it{i}$ filters, when available.  Links to the original images for each filter are published in the VOEvent Notice, and also provided on the DESAlert webpage\footnote{http://aao.gov.au/DESalert}.

In order to faciliate relative photometric measurements of the optical counterparts of GRBs, the VOEvent Notices provide magnitudes (and uncertainties) in each filter in a range of magnitudes (typically 16-26, depending on filter, with 1-2\% photometric accuracy) for several of the nearest stars within the search box centred on the GRB position.  Faint stars are far more likely to be positioned closest to any given GRB (and thus included in the XML notice), but information for all stars in the full region is stored in a similar format in the ancillary data available on the DESAlert website.

Likely of greatest interest to those desiring to follow up GRB observations are the DES pre-discovery catalogs of potential GRB host galaxies.  The VOEvent Notices provide positions, magnitudes (and uncertainties), photometric redshift information (when available), and chance alignment probabilities for the nearest galaxies within the search box centred on the GRB position.  Information for all galaxies in this same region is stored in a similar format in the ancillary data available on the DESAlert website.  For the subset of galaxies with no photometric redshift determination, we estimate the redshift ``on the fly'' with an empirical method based upon their relative $\it{gri}$ magnitudes \cite{Lopes}.  We only report physically realistic (i.e., positive) calculated values in the DESAlert data products (failures of the estimator are assigned a value of ``-9999'').  Because of the limited applicability of this empirical method, we stress that these values are approximations that will be supplanted by more precise determinations from the DES Data Management pipeline as they become available.

Host galaxies of long GRBs are expected to be precisely co-located with the GRB, while short GRBs may be several arcseconds distant from their original host.  If no galaxy is found within 10'' (for XRT-determined positions), the DESAlert VOEvent Notice instead reports the 10$\sigma$ galaxy detection limit derived \cite{Rykoff} from the Year 1 co-added observations -- $\it{g,r,i}$=23.4${\pm}$0.2, 23.1${\pm}$0.2, 22.5${\pm}$0.2, respectively, over the vast majority of the footprint (though not at the edges where there is less overlap among observations).  A galaxy undetected even down to these limits (especially as the Survey progresses and the limits are extended even fainter) may indicate a high-redshift GRB (and host) -- a particularly intriguing target for follow-up observations.  Meanwhile, the DESAlert database still includes all objects found within the full search area.

\section{Testing the DESAlert Algorithm}

In addition to being submitted to a ``code review'' by an experienced software engineer, each step of the DESAlert code outlined above has been individually tested, while end-to-end tests of the software show that all variations in expected VOEvent inputs are handled properly and output correct data products, even in rare and subtlely challenging cases (e.g., when a GRB is very close to RA=0.0, objects are returned with both RA$>$0 and RA$<$360).  Worth noting also for DESAlert users are ``edge'' cases where a GRB is near the border of the DES footprint (cf. Figure \ref{Fig2}).  Obviously, in this case DESAlert can only return objects observed by DES, so the output object list may not be a true reflection of all objects that would have been contained within the search box.  Therefore, a warning of potential incompleteness in the object catalogue is included with the VOEvent Notice.

By the conclusion of Year 1 in February 2014, DES had observed of order 2500 square degrees across the southern sky.  A variety of tests were performed with the DESAlert algorithm using both the Year 1 and Science Verification (``Year 0'') data \cite{Year1}.  First, we simulated VOEvent Notices with systematically-varying positions throughout the DES footprint -- though of course the real-time nature of DESAlert was not exercised in this way, we nevertheless could determine in a statistical way the expected impact of DESAlert.  Of the 45 simulated notices input to DESAlert, 23 were found to have archival DES data matching the locations, each of which had of order 500 objects (stars or galaxies) within 0.1 degrees of the nominal GRB position.  This yield is consistent with statistical expectations from a survey region that is 50$\%$ covered; with the data from Year 1 to Year 3 (and the start of Year 4) now populating the DES database, the future yield for DESAlert should be significantly higher.  Next, to test the real-time processing capabilities of DESAlert, a Swift VOEvent Notice of a newly-discovered GRB was received and processed, with the software taking approximately 1 minute to execute the full algorithm -- parsing the Notice, querying the relevant DES databases, building the VOEvent Notice, and ``submitting'' it (albeit to an internal recipient rather than the public VOEvent network).

\section{Results from the Application of DESAlert to Past Bursts}

To test the effectiveness of DESAlert specifically in finding GRB host galaxies, we compare positions of DESAlert galaxies with a selection of GRBs previously detected by Swift, along with a number of GRBs detected by BeppoSAX \cite{Beppo} and the satellites of the InterPlanetary Network \cite{IPN} such as HETE-II \cite{HETE}.  Many (though not all) of these bursts have spectroscopically-confirmed host galaxies drawn from The Optically Unbiased GRB Host (TOUGH) Survey \cite{TOUGH1, TOUGH3}, as well as GHostS, the GRB Host Studies \cite{GHostS}.  Of the 45 GRBs with XRT or other precise positions for which DESAlert searches returned nearby objects with photometric redshifts, sixteen were deemed sufficiently close (within 5'') to a galaxy to warrant more detailed examination.  These are GRBs 000210, 050219B, 060614A, 061007A, 071227A, 080514B, 080916A, 081109A, 090827A, 091018A, 110206A, 120701A, 140413A, 140928A, 151111A, and 160422A.

Of these sixteen, seven have spectroscopically confirmed redshifts, while nine do not.  The galaxy closest to GRB000210 (only 0.21'' away -- smaller than the pixel scale of DECam) has a photometric redshift derived by DES of z$_{phot}$ = 0.844 (a difference of 0.3\% from the spectroscopic redshift of z$_{spec}$=0.846).  Furthermore, though the filter response curves are different, DES photometry is consistent with previously reported observations of the host galaxy \cite{Christensen}.  Next, the galaxy closest to GRB081109A (2.93'' away -- well within the 4.8'' uncertainty reported by XRT) has a photometric redshift derived by DES of z$_{phot}$ = 0.906 (a difference of 8.0\% from the spectroscopic redshift of z$_{spec}$=0.979).  DES photometry is consistent with the upper limits on host magnitude from Swift UVOT \cite{8504}, the REM Telescope \cite{8501} and the Faulkes Telescope-South \cite{8508}, though all have significantly brighter (more than one magnitude) limits than the DES observations.  Third, the galaxy closest to GRB091018A (2.16'' away -- close, but outside the reported 0.6'' position uncertainty from XRT) has a photometric redshift derived by DES of z$_{phot}$=1.01 (a difference of 3.9\% from the spectroscopic redshift of z$_{spec}$=0.971).  Once again, DES photometry matches well with observations of the host galaxy by GROND, Gemini South/GMOS, and Faulkes Telescope-South \cite{Wiersema}.  The galaxies closest to the other four GRBs have photometric redshifts that do not reflect the spectroscopic redshifts as accurately (differences of 20\% to nearly 50\%).  See Table \ref{Tab0} for details of all of these candidate host galaxies, as well as the galaxies closest to the GRBs without spectroscopically confirmed redshifts.

\begin{sidewaystable*}[htbp]
\caption{Parameters of GRB-Possible Host Galaxy Matches, including GRB ID, Galaxy Position, Galaxy Magnitudes, Angular Separation from GRB, Chance Alignment Probability, Photometric Redshift, and Spectroscopic Redshift.}
%\begin{center}
\begin{tabular*}{\textwidth}{@{\extracolsep{\fill}}cccccccccccc@{}}
\hline\hline
GRB ID & Galaxy RA & Galaxy Dec & g Mag & r Mag & i Mag & z Mag & Y Mag & Angular Sep. & P$_{chance}$ & z$_{phot}$ & z$_{spec}$ \\
\hline
GRB000210& 29.815049 & -40.659132 & 24.1664 & 23.7162 & 23.3144	& 23.0766	& 21.0656	& 0.216778597 & 0.00011 & 0.844 & 0.846 \\
GRB060614A & 320.883702 & -53.026781& 23.0166 & 22.6119 & 22.428 & 22.2003 & 20.3721 & 0.493410174 & 0.00050 & 0.201 & 0.125 \\
GRB061007A & 46.331073	& -50.50043 & 24.2663 & 23.6688 & 23.5661 & 23.1845 & 22.6827 & 2.316367121 & 0.013 & 0.846	& 1.261 \\
GRB071227A & 58.129097 & -55.983558 & 22.1356 & 20.6839 & 20.1904 & 19.8608 & 19.8275 & 3.967900664 & 0.013 & 0.484 & 0.383 \\
GRB080916A & 336.275888 & -57.023007 & 23.2098 & 23.1742 & 23.0493 & 23.1942 & 21.4079 & 1.740304939 &	0.0063 & 0.505 & 0.689 \\
GRB081109A & 330.789946 & -54.711213 & 23.0523 & 22.6215 & 21.9645 & 21.9331 & 21.7905 & 2.932011596 & 0.018 & 0.906 & 0.979 \\
GRB091018A & 32.186478	& -57.548321 & 23.4613 & 23.3279 & 22.4687 & 22.2655 & 22.4532 & 2.160614912 & 0.0097 & 1.01 & 0.97 \\
\\
GRB050219B & 81.31725 & -57.758509 & 24.4297 & 24.0892 & 23.8385 & 22.547 & 19.9256 & 2.782734799 & 0.019 & 0.988 & N/A \\
GRB080514B & 322.844461 & -0.707721 & 24.982 & 23.4154 & 23.5333 & 22.1828 & 20.1415 & 2.150987197 & 0.012 & 1.159 & N/A \\
GRB090827A & 18.451531	& -50.896477 & 23.1477 & 22.8216 & 22.0605 & 21.4928 & 21.909 & 1.0392097 & 0.0022 & 0.904 &	N/A \\
GRB110206A & 92.33365	& -58.807486 & 22.9578 & 23.9749 & 23.2433 & 22.6352 & 21.7652 & 1.907782629 & 0.0031 & 1.05 & N/A \\
GRB140413A & 65.454482 & -51.182472 & 23.9548 & 23.7098 & 23.0644 & 22.5693 & 22.7619 & 2.491210405 & 0.013 & 0.964 & N/A \\
GRB140413A & 65.45538 & -51.182743 & 23.4559 & 23.1804 & 22.773 & 22.3432 & 22.3985 & 2.624765407 & 0.014 & 0.966 & N/A \\
GRB140928A & 43.69911 & -55.928973 & 23.803 & 23.1457 & 23.2778 & 22.9816 & 22.562& 1.271661464 & 0.0034 & 0.451	& N/A \\
GRB140928A & 43.699043	& -55.928131 & 23.1988 & 22.9256 & 22.7706 & 22.3542 & 21.5431 & 2.67615874 & 0.015 & 0.621 & N/A \\
GRB151111A & 56.844864 & -44.162079 & 23.1485 & 22.1382 & 21.4674 & 21.0164 & 20.7267 & 2.208525916 & 0.011 & 0.570 & N/A \\
GRB160422A & 42.09522 & -57.875097 & 23.0989 & 22.3213 & 21.9353 & 21.4508 & 18.3862 & 2.809451163 & 0.016 & 0.743 & N/A \\
\hline\hline
\end{tabular*}
%\end{center}
\label{Tab0}
\end{sidewaystable*}

\begin{figure}
\begin{center}
\includegraphics[width=\columnwidth]{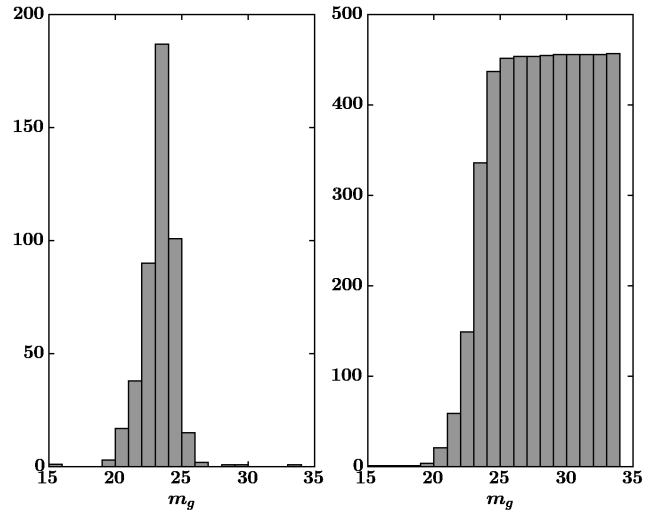}
\caption{(Left) Number of galaxies as a function of magnitude returned by DESAlert from 80'' x 80'' search boxes around 65 GRBs with BAT positions.  (Right) The cumulative number of galaxies equal to (or brighter than) each given magnitude.  The total number of objects classified as galaxies within these 416,000 square arcseconds is 454, resulting in (on average) approximately one galaxy per 910 square arcseconds.}\label{Fig6}
\end{center}
\end{figure}

To determine the general chance alignment of these galaxies with the GRB positions as determined by BAT, we start by finding the total number of galaxies in 80'' x 80'' searchboxes around a large subsample of all GRBs that were tested with DESAlert.  Of those searches that returned nearby objects (whether with photo-z's or not), there were 454 galaxies in the regions around 65 GRBs, totaling 416,000 square arcseconds -- approximately one galaxy per 910 square arcseconds -- see Figure \ref{Fig6}.  The probability of a chance alignment (P$_{chance}$) within a single 3.97'' x 3.97'' box (correspoding to the largest angular separation for any of the potential matches) is simply the ratio of the two areas; that is, P$_{chance}$=1.72\%.  Of course, more than a single search was performed; to determine the chance probability of no matches in 45 searches, we calculate P$_{45}$ = [1-P$_{chance}$]$^{45}$ = 0.458.  The chance probability of one or more matches for a single GRB, then, is 1 - P$_{45}$ = 0.542, while the chance probability of sixteen independent matches is P$_{16}$ = (0.542)$^{16}$ = 5.47 x 10$^{-5}$.  That is, the probability that at least one of the sixteen matches is the true host of the corresponding GRB is 99.9945\%.

\begin{figure}
\begin{center}
\includegraphics[width=\columnwidth]{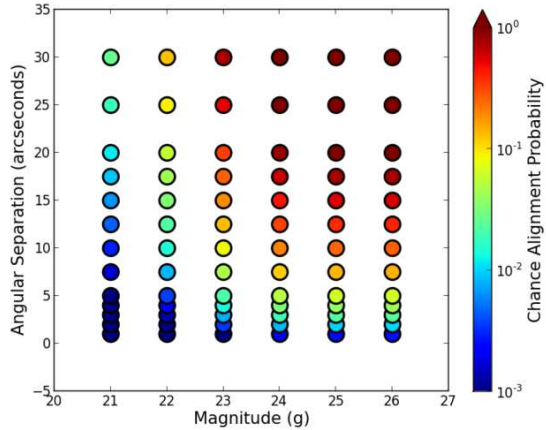}
\caption{Probability of chance alignment of galaxies within the specified angular separation from the nominal GRB position and brighter than the specified magnitude, derived from 30'' x 30'' search boxes around 38 GRBs.  The total number of objects classified as galaxies within these 34,200 square arcseconds is 89, resulting in (on average) approximately one galaxy per 384 square arcseconds.}\label{FigN}
\end{center}
\end{figure}

We also performed a more refined chance alignment analysis with a subsample of 38 GRBs with XRT positions, taking into account the individual galaxies' magnitudes and separations from the nominal GRB positions.  The total area searched was 34,200 sq. arcsec, within which 89 galaxies were found with magnitudes 20$<$g$<$26 and angular separations $<$30''.  We calculate that an angular separation of 0.216'' for a g$<$25$^{th}$ magnitude galaxy (cf. GRB00210) has a chance probability of only 0.011\%.  Similarly, angular separations of 2.93'' and 2.16'' for g$<$24$^{th}$ magnitude galaxies (cf. GRBs 081109A and 091018A) have a chance probability of 1.78\% and 0.968\%, respectively, though none of these probabilities factor in the entire ensemble of searches and matches.  Figure \ref{FigN} shows the cumulative chance alignment probability as a function of angular separation and galaxy magnitude.  For those wishing to make decisions regarding allocation of valuable follow-up observational resources, suitable cutoffs in probability space can be straightforwardly derived from this information -- specifically, we note that any bright (g$<$22$^{nd}$ magnitude) galaxy has a very low probability of chance alignment at any separation up to at least 20'', while any galaxy has a very low probability of being within 5'', regardless of magnitude.  We also note that these probabilities are, if anything, $\it{underestimates}$ of the significance of alignments, since the positions used in the analysis are, by construction, correlated with GRBs (and thus with the actual host galaxies), so any arbitrary point on the sky is $\it{less}$ likely to have a galaxy nearby than these positions.

%To determine the chance alignment of these galaxies with the GRB positions, we start by finding the total number of galaxies in 80'' x 80'' searchboxes around a large subsample of all GRBs that were tested with DESAlert.  Of those searches that returned nearby objects (whether with photo-z's or not), there were 454 galaxies in the regions around 65 GRBs, totaling 416000 square arcseconds -- approximately one galaxy per 910 square arcseconds -- see Figure \ref{Fig6}.  The probability of a chance alignment (P$_{chance}$) within a single 2.16'' x 2.16'' box (correspoding to the largest angular separation for any of the four potential matches) is simply the ratio of the two areas; that is, P$_{chance}$=0.509\%.  Of course, more than a single search was performed; to determine the chance probability of no matches in 55 searches, we calculate P$_{55}$ = [1-P$_{chance}$]$^{55}$ = 0.759.  The chance probability of one or more matches for a single GRB, then, is 1 - P$_{55}$ = 0.214, while the chance probability of four independent matches is P$_{4}$ = (0.214)$^{4}$ = 0.00336.  That is, the probability that at least one of the four matches is the true host of the corresponding GRB is 99.66\%.

Given the low probabilities associated with the observed galaxy alignments, the correspondence between the spectroscopic and photometric redshifts, and the similarity of the DES and non-DES photometry, we claim that DESAlert has successfully found the (previously-known) host for GRB000210, and has with high probability found the (previously-known) hosts of GRB081109A and GRB091018A.  Though in the absence of spectroscopic redshift information we cannot say with certainty that the additional galaxy matches are indeed the hosts of GRBs 050219B, 080514B, 090827A, 110206A, 120701A, 140413A, 140928A, 151111A, or 160422A, the low probabilities of chance alignment alone provide intriguing evidence pointing in that direction.  Most importantly, we show that DESAlert can provide ideal candidates for real-time follow-up observations that could confirm or refute the status of proposed GRB host galaxies.

After completing the search of archival bursts, we tested the DESAlert algorithm with bursts detected by Swift during the Year 1 observations of DES.  In Appendices A and B, we present an example of the data products related to a single burst (GRB131105A) that are provided to the community by DESAlert -- specifically, the VOEvent XML Notice and Finder Image, respectively.  Table \ref{Tab1} shows a subset of the relevant data for selected stars and galaxies near this GRB as well.  Table entries include object position, $\it{grizY}$ magnitudes, the $\it{spread}$$\_$$\it{model}$ value, object separation from the nominal GRB positon (in arcseconds), object classification, chance alignment probability, photometric redshift, and object number as defined internally by DESAlert (where the same number indicates additional observational epochs of the same object).  $\it{Spread}$$\_$$\it{model}$ is a neural-network based star/galaxy classifier within SourceExtractor; values greater than 0.003 correspond to galaxies \cite{DESDM}.  We calculate the mean of the $\it{spread}$$\_$$\it{model}$ values for all filters, with each weighted by the inverse square of the uncertainty in the observations for that filter, yielding a value that relies most heavily upon the most accurate observation, but still incorporates all available data.  As with other GRBs we tested, the photometric redshifts are calculated following the method described above in Section \ref{sec:products} -- for those observations with physically realistic values (i.e., z$_{phot}$ $>$ 0), the values derived from repeated observations of the same object are broadly consistent (generally within $\pm$10-20$\%$ of one another).  All stars and galaxies in the region around the GRB are stored in a similar fashion in the ancillary data available on the DESAlert website.

\begin{sidewaystable*}[htbp]
\caption{Parameters of Selected Galaxies and Stars near GRB131105A, including Position, Magnitudes, Star/Galaxy Classification ($\it{Spread}$$\_$$\it{Model}$) Value, Angular Separation from GRB, Star/Galaxy Classification, Chance Alignment Probability, Photometric Redshift, and DESAlert Object ID (useful particularly for identifying duplicate observations of the same object).}
%\begin{center}
\begin{tabular*}{\textwidth}{@{\extracolsep{\fill}}ccccccccccccc@{}}
\hline\hline
RA & DEC & g Mag & r Mag & i Mag & z Mag & Y Mag & Spread$\_$Model & Angular Sep. & Classification & P$_{chance}$ & z$_{phot}$ & Obj. ID \\
\hline
70.968172	& -62.99296 & 22.503	& 22.2323	& 21.8513	& 21.4339	& 20.9653	&0.015399104	& 8.472173	 & 'galaxy'	& 0.065 & 0.703 &	0 \\
70.966681	& -62.99750 & 22.966 & 22.1018	& 21.873 & 21.5775 & 21.8765 & 0.010310614 & 8.748330364 & `galaxy' & 0.10 & 0.4366 & 1 \\
70.967236	& -62.99211 & 22.1341 & 22.3432 & 21.0289 & 21.4836 & 21.3055 & 0.00484939 & 11.11854915 & `galaxy' & 0.12 & 1.073 & 2 \\
70.968191 & -62.998427 & 22.9998 & 22.248 & 21.3831 & 21.301& 21.0583 & 0.015187986 & 11.97919136 & `galaxy' & 0.40 & 0.797 & 3 \\
70.96651 & -62.99142 & 23.2673 & 22.5724 & 21.8852 & 21.2592 & 20.705 & 0.012680891 & 13.96178212 & `galaxy' & 0.70 & 0.923 & 4 \\
70.970161 & -62.991469 & 24.3125 & 23.705 & 23.152 & 22.4144 & 22.7957 &  0.006773822 & 16.63765696 & `galaxy' & 0.98 & 1.037 & 5 \\
70.972785 & -62.996331 & 24.1598 & 23.3784 & 22.9457 & 22.5508 & 30.5481 &  0.003969985 & 19.74596095 & `galaxy' & $\approx$1 & 0.958 & 6 \\
70.96423 & -62.988984 & 22.4958 & 20.6742 & 19.8973 & 19.4721 & 19.3179 & 0.012232375 &	 	25.1202975 & `galaxy' & 0.54 & 0.550 & 7 \\
70.97038 & -63.00157 & 23.6146	 & 23.2928 & 22.3022 & 21.915 & 21.6459 & 0.007732379 & 25.31954502 & `galaxy' & $\approx$1 & 0.984 & 8 \\
70.961336 & -62.990123 & 24.2874 & 23.4033 & 22.8116 & 22.9804 & 20.8511 & 0.008236298 & 28.50362263 & `galaxy' & $\approx$1 & 0.585 & 9 \\
70.960276 & -62.998921 & 23.579 & 23.6619 & 22.7834 & 21.8106 & 22.315 & 0.012734838 & 29.01454768 & `galaxy' & $\approx$1 & 1.094 & 10 \\
70.974674 & -62.99898 & 24.4774 & 23.2527 & 23.2273 & 22.4959 & 23.2027 & 0.008000411 & 29.46388677 & galaxy' & $\approx$1 & 0.361 & 11 \\ 
\\
70.967696 & -62.99528 & 24.4594 & 24.5978 & 24.2158 & 21.2049 & 21.7938 & 0.000773172 & 	1.045091843 & `star' & N/A & N/A & 12 \\
70.965281 & -62.994298 & 22.6208 & 22.4381 & 22.2976 & 22.0604 & 22.0944 & -0.000418506 &	8.343138834 & `star' & N/A & N/A & 13 \\
70.968682 & -62.992487 & 21.9626 & 21.6927 & 21.455 & 21.528 & 21.0872 & 0.001480552 & 10.73914032 & `star' & N/A & N/A & 14 \\
70.963219 & -62.995329 & 24.6899 & 24.8741 & 23.7952 & 22.7461 & 22.7624 & 0.001944441 &	15.13187619 & `star' & N/A & N/A & 15 \\
70.9617 & -62.996758 & 25.0523	 & 24.2477 & 24.0528 & 23.8909 & 22.4848 & -0.010672662 & 	21.35167982 & `star' & N/A & N/A & 16 \\
70.971987 & -63.000353 & 24.427 & 24.0121 & 24.0557 & 23.4534 & 21.4052 & -0.001584357 & 	24.81495903 & `star' & N/A & N/A & 17 \\
70.973996 & -62.997374 & 23.608 & 23.4864 & 23.2615 & 23.013 & 22.5031  & -0.001571138 & 	24.94507308 & 	`star' & N/A & N/A & 18 \\
70.962769 & -62.989363 & 22.964 & 21.2918 & 21.9828 & 20.646 & 20.2515 & -0.000343895 & 	26.84010173 & `star' & N/A & N/A & 19 \\
70.966285 & -63.003414 & 24.2086 & 23.7094 & 23.2906 & 23.0829 & 21.2759 & -0.002716397 & 	29.8870259 & `star' & N/A & N/A & 20 \\
\hline\hline
\end{tabular*}
%\end{center}
\label{Tab1}
\end{sidewaystable*}

\section{Future Development of DESAlert}
\label{sec:future}

Though DESAlert is fully functional and has made several confirmed (and proposed) matches to GRB hosts, we will continue to improve the functionality of DESAlert through future code releases.  Dubbed DESAlert++, the next iteration of the DESAlert algorithm will incorporate several new aspects.  In particular, we are investigating alternate search methods to including transients with larger position uncertainties, such as Fermi-detected GRBs with position uncertainties close to 1 degree \cite{Fermi}.  One particularly promising category of transients for inclusion in DESAlert++ is Fast Radio Bursts \cite{FRB}.  Though their positions are not well known (for example, the beam size of the Parkes radio telescope that has detected numerous FRBs is 14' in diameter), their redshift can be significantly constrained -- for example, \cite{Petroff} have determined that z$<$0.5 for FRB 140514.  Despite the probability of many galaxies being observed by DES in a search box more than 10 arcminutes across, photometric redshift matches could positively identify optical counterparts (or hosts) of FRBs, or at the very least facilitate statistical approaches to identifying likely sources (similar to the method used above to identify multiple potential new GRB host galaxies).

Next, multi-messenger transient detections (e.g., gravitational wave observations) stemming from the Astrophysical Multimessenger Observatory Network \cite{AMON} provide yet another potentially interesting avenue of further development of DESAlert.  Finally, we are exploring the possibility of including resolved stellar transients such as disappearing red supergiants \cite{Kochanek} or galactic stellar flares --- in such cases, nearby stars (rather than galaxies) become the interesting targets for which the transient follow-up community can utilize the data products released by DESAlert.

\section{Conclusions}
\label{sec:conclusions}

DESAlert is an algorithm implemented in Python and SQL to receive automated notices of GRB parameters from VOEvent Notification triggers, and to provide the astronomical community with finder images as well as catalogs of nearby stars and galaxies (with relevant quantities such as positions, magnitudes, and photometric redshifts).  The details of the algorithm have been laid out in the previous Sections, and a sample image and catalog shown to familiarise readers with the DESAlert data products.  Based on tests using historical GRBs, we confirm the effectiveness of DESAlert by finding multiple previously-discovered GRB hosts (matching in both position and redshift, and consistent with independent photometric observations), and further by detecting several potential GRB hosts (matching in position, but with no spectroscopic redshift to confirm the DES-derived photometric redshift).

During the five-year lifetime of the DES, we expect to provide data products for of order 100 Swift-detected GRBs -- though the DESAlert system is expected to function well beyond the formal lifetime of the Dark Energy Survey itself.  Extensions to DESAlert (dubbed DESAlert++) are currently being explored based upon other GRB-detecting instruments such as Fermi (or based on the incorporation of other transients such as FRBs into the DESAlert algorithm).  To estimate the annual rate of GRBs from all sources likely to be within the DES footprint, we searched the SkyAlert database for unique GRB positions from all sources from one year; of the 230 bursts in the database, 34 (roughly 3 per month) fall within the DES footprint and could potentially trigger DESAlert++.

We encourage all observers interested in follow-up observations of transient astrophysical phenomena such as GRBs to avail themselves of the VOEvent Notices and web-based data archive provided by DESAlert.

\begin{acknowledgements}
The authors thank Simon O'Toole for assistance in setup and maintenance of the DESAlert Data Archive at the Australian Astronomical Observatory, as well as Nuria P.F. Lorente for a comprehensive review of the DESAlert code, and Dr. Tayyaba Zafar for providing archival GRB data for testing DESAlert.  We also thank the anonymous referee for many insightful comments that prompted numerous improvements in the manuscript and the functionality of the DESAlert code.  KK is grateful to the Astronomical Society of Australia for convening an Early Career Researcher Writing Workshop, during which a significant portion of this text was written.  The authors also recognize the Macquarie University PACE program and Professor Mike Ireland for facilitating the undergraduate research opportunities, as well as Raymond Lam for early efforts toward DESAlert.  This research has made use of the GHostS database (www.grbhosts.org), which is partly funded by Spitzer/NASA grant RSA Agreement No. 1287913, and has also made use of the SIMBAD database, operated at CDS, Strasbourg, France.  We are grateful for the extraordinary contributions of our CTIO colleagues and the DECam Construction, Commissioning and Science Verification teams in achieving the excellent instrument and telescope conditions that have made this work possible.  The success of this project also relies critically on the expertise and dedication of the DES Data Management group.

This paper has gone through internal review by the DES collaboration.  Funding for the DES Projects has been provided by the U.S. Department of Energy, the U.S. National Science Foundation, the Ministry of Science and Education of Spain, the Science and Technology Facilities Council of the United Kingdom, the Higher Education Funding Council for England, the National Center for Supercomputing Applications at the University of Illinois at Urbana-Champaign, the Kavli Institute of Cosmological Physics at the University of Chicago, the Center for Cosmology and Astro-Particle Physics at the Ohio State University, the Mitchell Institute for Fundamental Physics and Astronomy at Texas A\&M University, Financiadora de Estudos e Projetos, Funda{\c c}{\~a}o Carlos Chagas Filho de Amparo {\`a} Pesquisa do Estado do Rio de Janeiro, Conselho Nacional de Desenvolvimento Cient{\'i}fico e Tecnol{\'o}gico and the Minist{\'e}rio da Ci{\^e}ncia e Tecnologia, the Deutsche Forschungsgemeinschaft and the Collaborating Institutions in the Dark Energy Survey. 

The DES data management system is supported by the National Science Foundation under Grant Number AST-1138766.
The DES participants from Spanish institutions are partially supported by MINECO under grants AYA2012-39559, ESP2013-48274, FPA2013-47986, and Centro de Excelencia Severo Ochoa SEV-2012-0234, some of which include ERDF funds from the European Union.

The Collaborating Institutions are Argonne National Laboratory, the University of California at Santa Cruz, the University of Cambridge, Centro de Investigaciones Energeticas, Medioambientales y Tecnologicas-Madrid, the University of Chicago, University College London, the DES-Brazil Consortium, the Eidgen{\"o}ssische Technische Hochschule (ETH) Z{\"u}rich, Fermi National Accelerator Laboratory, the University of Edinburgh, the University of Illinois at Urbana-Champaign, the Institut de Ciencies de l'Espai (IEEC/CSIC), the Institut de Fisica d'Altes Energies, Lawrence Berkeley National Laboratory, the Ludwig-Maximilians Universit{\"a}t and the associated Excellence Cluster Universe, the University of Michigan, the National Optical Astronomy Observatory, the University of Nottingham, The Ohio State University, the University of Pennsylvania, the University of Portsmouth, SLAC National Accelerator Laboratory, Stanford University, the University of Sussex, and Texas A\&M University.
\end{acknowledgements}

\onecolumn{
\begin{appendix}
\section{Appendix A: Sample DESAlert VOEvent Notice}
\label{sec:appendA}
\begin{verbatim}
<?xml version="1.0" ?>
<voe:VOEvent xmlns:xsi="http://www.w3.org/2001/XMLSchema-instance"
xmlns:voe="http://www.ivoa.net/xml/VOEvent/v2.0"
xsi:schemaLocation="http://www.ivoa.net/xml/VOEvent/v2.0 http://www.ivoa.net/xml/VOEvent/VOEvent-v2.0.xsd"
 version="2.0" role="test" ivorn="ivo://DESAlert.AAO/DESAlert#2014-08-16T15:50:52.00_16885">
    <Who>
        <Author>
            <shortName>DESAlert</shortName>
            <contactName>Kyler Kuehn</contactName>
            <contactEmail>kyler.kuehn@aao.gov.au</contactEmail>
        </Author>
    </Who>
    <What>
        <Param name="Event_TJD" dataType="string" value="16885" ucd="time" unit="days"/>
        <Param name="TrigID" dataType="string" value="1539486996" ucd="meta.id"/>
        <Group name="Full_Data_Set">
            <Param name="collaboration" dataType="string" value="collabURL" ucd="meta.ref.url"/>
            <Param name="finder_chart" dataType="string" value="finderURL" ucd="meta.ref.url"/>
            <Param name="data_tables" dataType="string" value="dataURL" ucd="meta.ref.url"/>
        </Group>
        <Table name="Nearest Galaxies">
            <Description> Positions of the nearest galaxies. </Description>
            <Field dataType="string" name="Type"/>
            <Field dataType="float" name="RA"/>
            <Field dataType="float" name="DEC"/>
            <Field dataType="float" name="Mag_g"/>
            <Field dataType="float" name="Magerr_g"/>
            <Field dataType="float" name="Mag_r"/>
            <Field dataType="float" name="Magerr_r"/>
            <Field dataType="float" name="Mag_i"/>
            <Field dataType="float" name="Magerr_i"/>
            <Field dataType="float" name="Mag_z"/>
            <Field dataType="float" name="Magerr_z"/>
            <Field dataType="float" name="Mag_y"/>
            <Field dataType="float" name="Magerr_y"/>
            <Field dataType="float" name="photo_z"/>
            <Field dataType="float" name="ang_sep"/>
            <Field dataType="float" name="P_chance"/>
            <Data>
                <TR>
                    <TD>Galaxy</TD>
                    <TD>70.968172</TD>
                    <TD>-62.99296</TD>
                    <TD>22.503</TD>
                    <TD>0.0881</TD>
                    <TD>22.2323</TD>
                    <TD>0.0672</TD>
                    <TD>21.8513</TD>
                    <TD>0.082</TD>
                    <TD>21.4339</TD>
                    <TD>0.1492</TD>
                    <TD>20.9653</TD>
                    <TD>0.429</TD>
                    <TD>0.7035380006</TD>
                    <TD>8.4721730294</TD>
                    <TD>0.0608</TD>
                </TR>
            Additional entries removed from sample Notice...
            </Data>
        </Table>
        <Table name="Nearest Stars">
            <Description> Positions of the nearest stars. </Description>
            <Field dataType="string" name="Type"/>
            <Field dataType="float" name="RA"/>
            <Field dataType="float" name="DEC"/>
            <Field dataType="float" name="Mag_g"/>
            <Field dataType="float" name="Magerr_g"/>
            <Field dataType="float" name="Mag_r"/>
            <Field dataType="float" name="Magerr_r"/>
            <Field dataType="float" name="Mag_i"/>
            <Field dataType="float" name="Magerr_i"/>
            <Field dataType="float" name="Mag_z"/>
            <Field dataType="float" name="Magerr_z"/>
            <Field dataType="float" name="Mag_y"/>
            <Field dataType="float" name="Magerr_y"/>
            <Field dataType="float" name="photo_z"/>
            <Field dataType="float" name="ang_sep"/>
            <Data>
                <TR>
                    <TD>Star</TD>
                    <TD>70.967696</TD>
                    <TD>-62.99528</TD>
                    <TD>24.4594</TD>
                    <TD>0.1771</TD>
                    <TD>24.5978</TD>
                    <TD>0.252</TD>
                    <TD>24.2158</TD>
                    <TD>0.2083</TD>
                    <TD>21.2049</TD>
                    <TD>0.5427</TD>
                    <TD>21.7938</TD>
                    <TD>0.9756</TD>
                    <TD>1.1668900251</TD>
                    <TD>1.04509184285</TD>
                </TR>
            Additional entries removed from sample Notice...
            </Data>
        </Table>
    </What>
    <WhereWhen>
        <ObsDataLocation>
            <ObservatoryLocation id="GEOLUN"/>
            <ObservationLocation>
                <AstroCoordSystem id="UTC-FK5-GEO"/>
                <AstroCoords coord_system_id="UTC-FK5-GEO">
                    <Time unit="s">
                        <TimeInstant>
                            <ISOTime>2014-08-16T15:50:52.00</ISOTime>
                        </TimeInstant>
                    </Time>
                    <Position2D unit="deg">
                        <Name1>RA</Name1>
                        <Name2>Dec</Name2>
                        <Value2>
                            <C1>70.967420</C1>
                            <C2>-62.995190</C2>
                        </Value2>
                        <Error2Radius>0.066600</Error2Radius>
                    </Position2D>
                </AstroCoords>
            </ObservationLocation>
        </ObsDataLocation>
        <Description>The RA,Dec coordinates are of the type: source_object.</Description>
    </WhereWhen>
    <How>
        <Description>This program uses the DES database to compile a list of nearby objects.</Description>
        <Description>DES uses the Blanco 4m telescope on Cerro Tololo, equipped with the DECam</Description>
        <Reference type="url" uri="http://gcn.gsfc.nasa.gov/swift.html"/>
        <Reference type="url" uri="http://lib.skyalert.org/VOEventLib/"/>
        <Reference type="url" uri="https://www.darkenergysurvey.org/DECam/DECam_add_tech.shtml"/>
    </How>
    <Citations>
        <EventIVORN cite="followup">ivo://nasa.gsfc.gcn/SWIFT#BAT_SubSubThresh_Pos_1539486996-439</EventIVORN>
        <Description>Telescope used: Swift Satellite, XRT Instrument</Description>
    </Citations>
    <Description>GRB real-time followup with additional objects</Description>
</voe:VOEvent>
\end{verbatim}
\end{appendix}

\begin{appendix}
\section{Appendix B: Example Finder Image}
\label{sec:appendB}
\begin{figure}
\begin{center}
\includegraphics[width=\columnwidth]{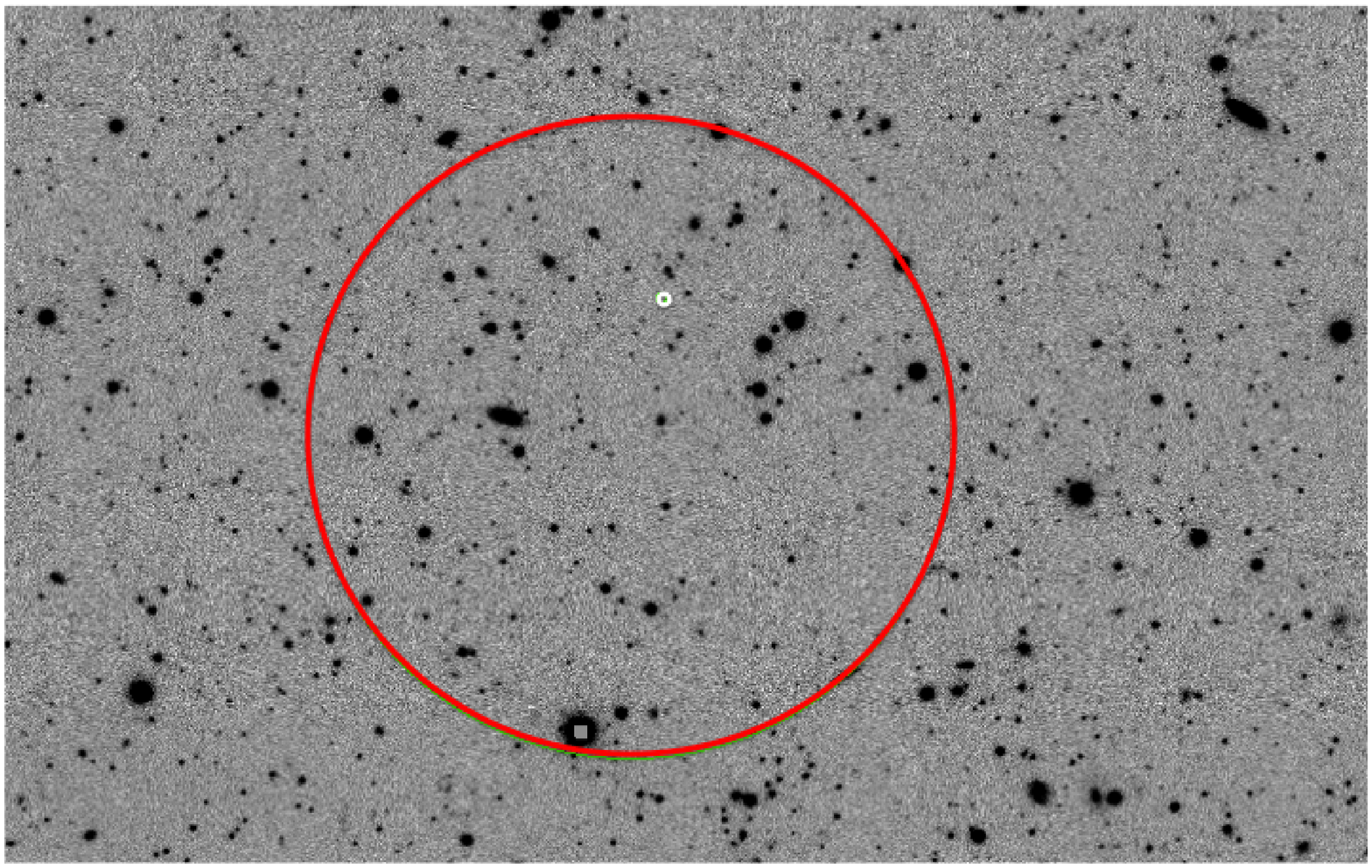}
% changed from f6.eps!
%\includegraphics{fpo.eps}
\caption{This is an example of (a zoomed-in segment of) a GRB Finder Image.  The initial Swift-BAT 90$\%$ error circle with 1.5' radius is shown (large red circle), as is the refined Swift-XRT 90$\%$ error circle with 1.4'' radius (small white circle).}\label{Fig7}
\end{center}
\end{figure}
\end{appendix}

% UNCOMMENT THE LINES BELOW IF YOU WISH TO USE BIBTEX
%\bibliographystyle{apj}
%\bibliography{yourbibfile}

\end{document}